\documentclass[twocolumn,showpacs]{revtex4}

\usepackage{graphicx}
\usepackage{dcolumn}
\usepackage{amsmath}
\begin{document}

\title{Resistivity and magnetization of
U(Ru$_{1-x}$Rh$_x$)$_2$Si$_2$, $x=$~2, 2.5, 3 and 4~\% in magnetic fields of up to 45~T}
\author{K.~H.~Kim$^{1,*}$, N. Harrison$^1$, H.~Amitsuka$^2$,
G.~A.~Jorge$^1$, M.~Jaime$^1$, and J. A. Mydosh$^{3,4}$}
\address{$^{1}$National High Magnetic Field Laboratory, Los Alamos National
Laboratory, MS E536, Los Alamos, NM 87545\\
$^{2}$Graduate School of Science, Hokkaido University, N10W8 Sapporo 060-0810,
Japan\\
$^{3}$Kamerlingh Onnes Laboratory, Leiden University, 2300RA Leiden, The
Netherlands. \\
$^{4}$Max-Planck-Institut f\"{u}r Chemische Physik fester Stoffe,
01187 Dresden,
Germany.}

\begin{abstract}
This posting contains raw data used for extracting features of the phase diagram and quantum criticality versus magnetic field $B$, temperature $T$ and Rh-doping $x$ that will soon appear in Physical Review Letters, 2004. All of the raw resistivity data is shown while only a selection of the magnetization data is shown for which a sample-out background subtraction was made.
\end{abstract}

\maketitle

\begin{figure}
\caption{Resistivity data for $x=$~2, 2.5, 3 and 4~\% obtained on sweeping the hybrid magnetic field up (except where indicated) at constant temperature. The temperature was stabilized using a capacitance thermometer. For $x=$~2, 2.5 and 3~\%, $T=$~0.3, 1.5, 1.6, 2.2, 3, 3 (down sweep), 4, 5.7, 6.8, 8, 8.7, 10, 12.5, 15, 16.5, 20.8, 29 and 35.5~K. For $x=$~4~\%, $T=$~0.3, 1.5. 2.2, 3, 4, 5.7 (down sweep), 6.8, 8, 8.7, 10, 12.5, 16.5, 20.8, 24, 33.5 and 35.5~K. In the case of $x=$~2 and $x=$~4~\%, the resistivity has been precisely calibrated. In the case of $x=$~2.5 and $x=$~3~\%, the data is renormalized.}
\label{fig1}
\end{figure}

\begin{figure}
\caption{Resistivity data for $x=$~2~\% measured at constant magnetic field in the hybrid magnet in Tallahassee, Florida. Field values are indicated.}
\label{fig2}
\end{figure}

\begin{figure}
\caption{Resistivity data for $x=$~2.5~\% measured at constant magnetic field in the hybrid magnet. Field values are indicated.}
\label{fig3}
\end{figure}

\begin{figure}
\caption{Resistivity data for $x=$~3~\% measured at constant magnetic field in the hybrid magnet. Field values are indicated.}
\label{fig4}
\end{figure}

\begin{figure}
\caption{Resistivity data for $x=$~4~\% measured at constant magnetic field in the hybrid magnet. Field values are indicated.}
\label{fig5}
\end{figure}

\begin{figure}
\caption{Magnetization data (arbitrary units) for $x=$~2~\% measured in the 50~T mid-pulse magnet at Los Alamos National Laboratory. An extraction magnetometer was used which enables recorded data without the sample present to be subtracted. Additional data without background subtraction, that was used to obtain some of the sharper phase transition features published in Physical Review Letters is not shown. Black, red, green, blue, cyan, magenta, yellow, dark yellow and navy blue curves correspond to $T=$~0.5, 4.4, 5.5, 6.5, 7, 7.5, 10, 14 and 18~K respectively.}
\label{fig6}
\end{figure}

\begin{figure}
\caption{Magnetization data (arbitrary units) for $x=$~2.5~\% measured in the 50~T mid-pulse magnet at Los Alamos National Laboratory. An extraction magnetometer was used which enables recorded data without the sample present to be subtracted. Additional data without background subtraction, that was used to obtain some of the sharper phase transition features published in Physical Review Letters is not shown. Black, red, green, blue, cyan and magenta curves correspond to $T=$~1.5, 6.25, 7, 10, 14 and 18~K respectively.}
\label{fig7}
\end{figure}

\begin{figure}
\caption{Magnetization data (arbitrary units) for $x=$~3~\% measured in the 50~T mid-pulse magnet at Los Alamos National Laboratory. An extraction magnetometer was used which enables recorded data without the sample present to be subtracted. Additional data without background subtraction, that was used to obtain some of the sharper phase transition features published in Physical Review Letters is not shown. Black, red, green, blue, cyan and magenta curves correspond to $T=$~0.5, 7.5, 8.5, 10, 14 and 18~K respectively.}
\label{fig8}
\end{figure}

\begin{figure}
\caption{Magnetization data (arbitrary units) for $x=$~4~\% measured in the 50~T mid-pulse magnet at Los Alamos National Laboratory. An extraction magnetometer was used which enables recorded data without the sample present to be subtracted. Additional data without background subtraction, that was used to obtain some of the sharper phase transition features published in Physical Review Letters is not shown. }
\label{fig9}
\end{figure}


\begin{thebibliography}{99}
\bibitem[$*$]{khkim}present address: CSCMR \& School of Physics, Seoul National
University, Seoul 151-742, Korea.


\end{thebibliography}
\end{document}